%% file: paper.tex
\newif\ifpdf
\ifx\pdfoutput\undefined
        \pdffalse 
\else
        \pdfoutput=1
        \pdftrue
\fi

\documentclass[fleqn,twoside]{article}
\usepackage{espcrc2}
\ifpdf
	\usepackage[pdftex]{graphicx}
	\DeclareGraphicsExtensions{.jpg,.pdf,.mps,.png}
\else
	\usepackage[xdvi,dvips]{graphicx}
\fi

\input Macros.tex

\title{Properties of charmonium in lattice QCD with ${2+1}$
flavors of improved staggered sea quarks
}

\author{Massimo~di~Pierro\address[DePaul]{School of Computer 
Science, Telecommunications, and Information Systems, DePaul University, Chicago, IL 60604},
Aida~X.~El-Khadra\address[UIUC]{
Department of Physics, University of Illinois, Urbana, IL 61801},
Steven~Gottlieb\address[IU]{
Department of Physics, Indiana University, Bloomington, IN 47405},
Andreas~S.~Kronfeld\address[FNAL]{
Fermi National Accelerator Laboratory, P.O. Box 500, Batavia, IL 60510}, 
Paul~B.~Mackenzie\addressmark[FNAL],
Damian~P.~Menscher\addressmark[UIUC], Mehmet~B.~Oktay\addressmark[UIUC],
Masataka Okamoto\addressmark[FNAL],
and James~N.~Simone\addressmark[FNAL]\thanks{Talk presented by J. Simone,
simone@fnal.gov}
}

\newcommand{\FIGccSpectrum}{
\begin{figure}[htb]
\ifpdf
   \includegraphics[clip=true,width=0.95\columnwidth]{chiral-cc_spectrum.png}
\else
   \includegraphics[clip=true,width=0.95\columnwidth]{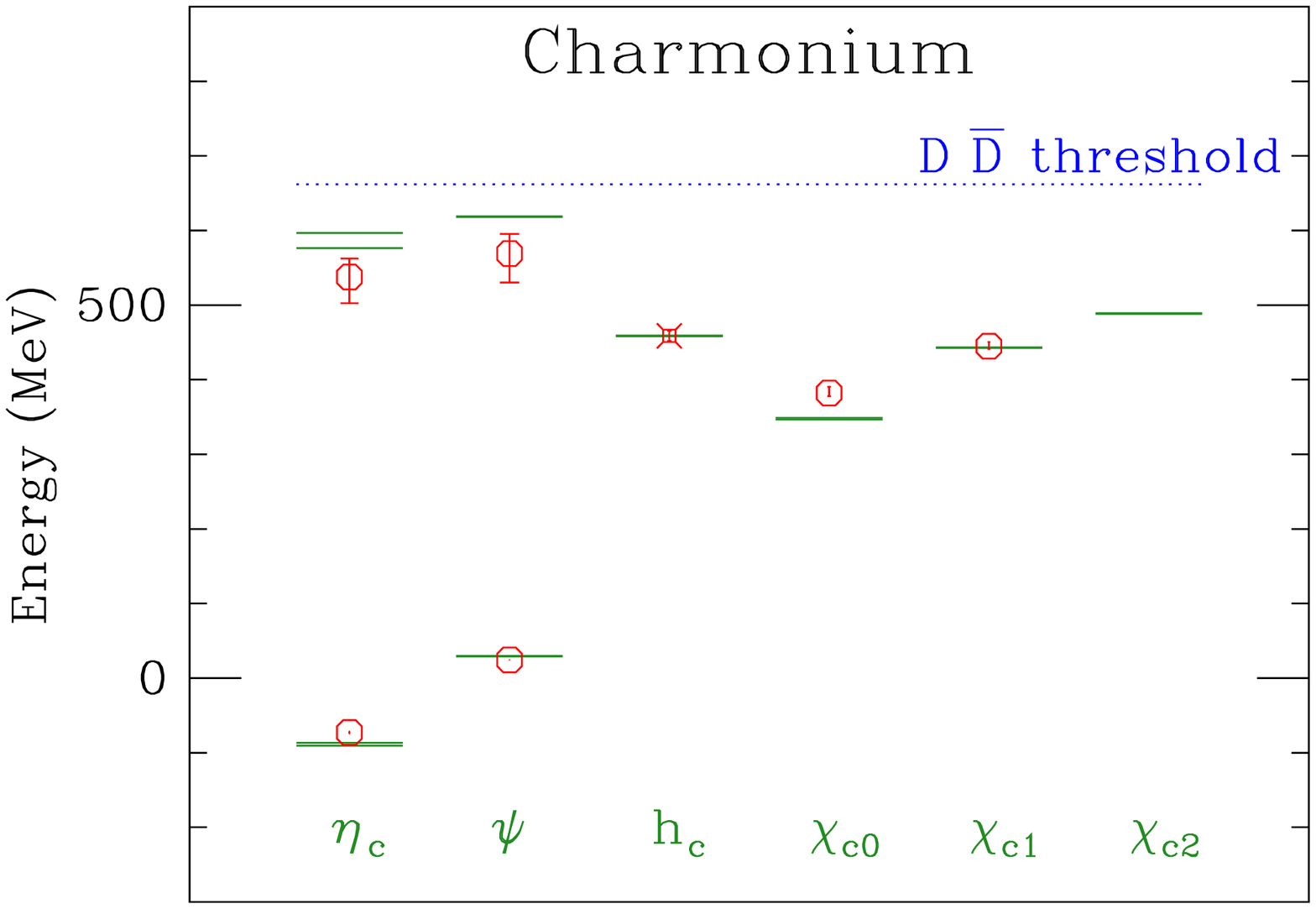}
\fi
\vspace{-1.0ex}
\caption{The charmonium spectrum in $2+1$ flavor lattice QCD
after extrapolation to the physical sea quark masses.
The lattice spacing is $a\approx1/8\,\fm$.
No lattice result for the $\chi_{c2}(\OneP)$ splitting is shown
since that state was not computed in this study.
The dotted indicates the $D\,\bar{D}$ threshold energy.
}
\label{fig:ccSpectrum}
\end{figure}}

\newcommand{\FIGhyperfine}{
\begin{figure}[htb]
\ifpdf
   \includegraphics[clip=true,width=0.95\columnwidth]{chiral-psi1S-eta_c1S.png}
\else
   \includegraphics[clip=true,width=0.95\columnwidth]{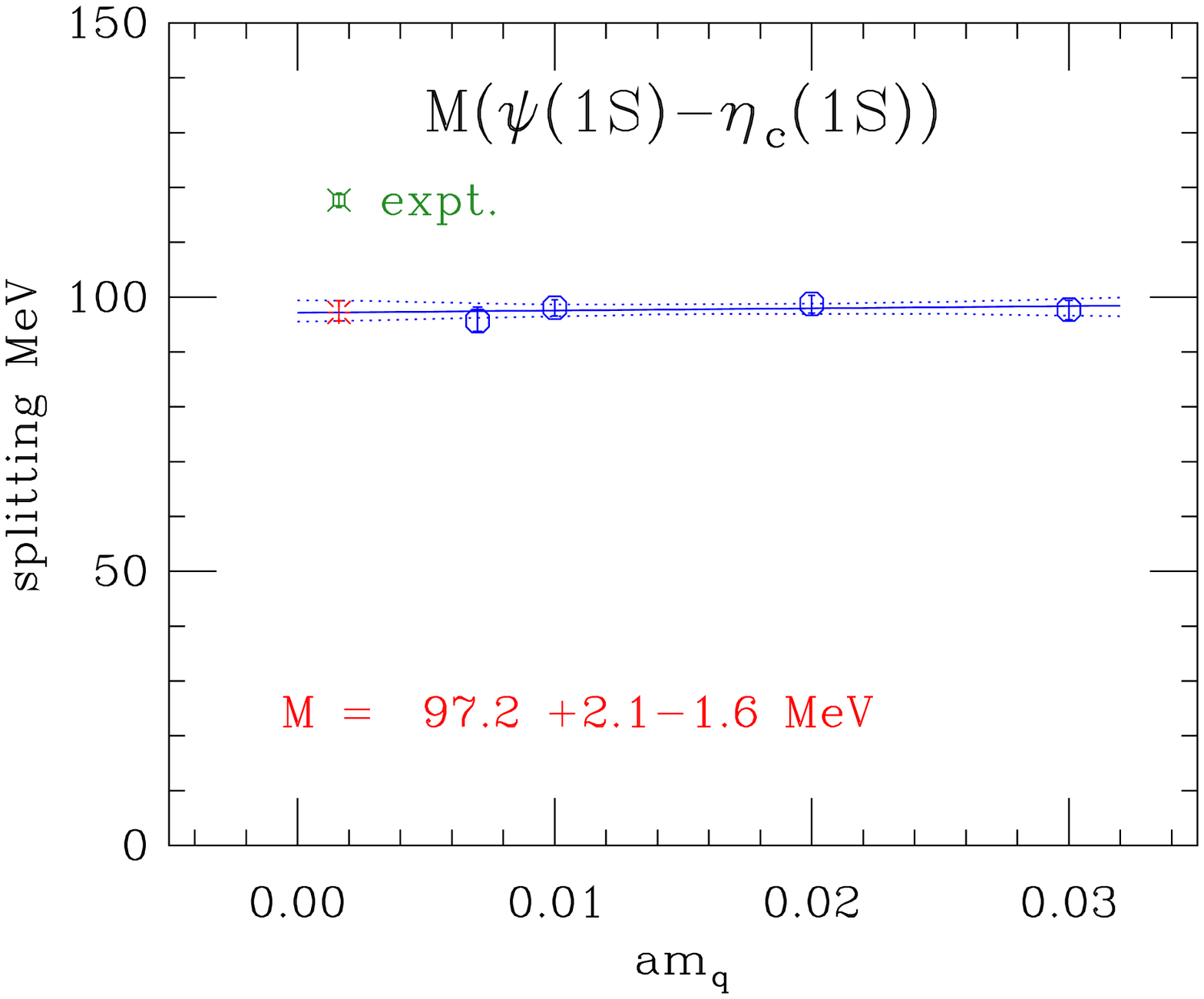}
\fi
\vspace{-5ex}
\caption{The charmonium $\OneS$ hyperfine splitting as a function
of the mass of the two light sea quarks. The solid and
dotted lines show the linear
extrapolation with 68\% confidence bounds. The extrapolated value
is denoted by the burst symbol. The experimental value lies above
the lattice determination.}
\label{fig:hyperfine}
\end{figure}}

\newcommand{\FIGleptonic}{
\begin{figure}[htb]
\ifpdf
   \includegraphics[clip=true,width=0.95\columnwidth]{chiral-psi-leptonic.png}
\else
   \includegraphics[clip=true,width=0.95\columnwidth]{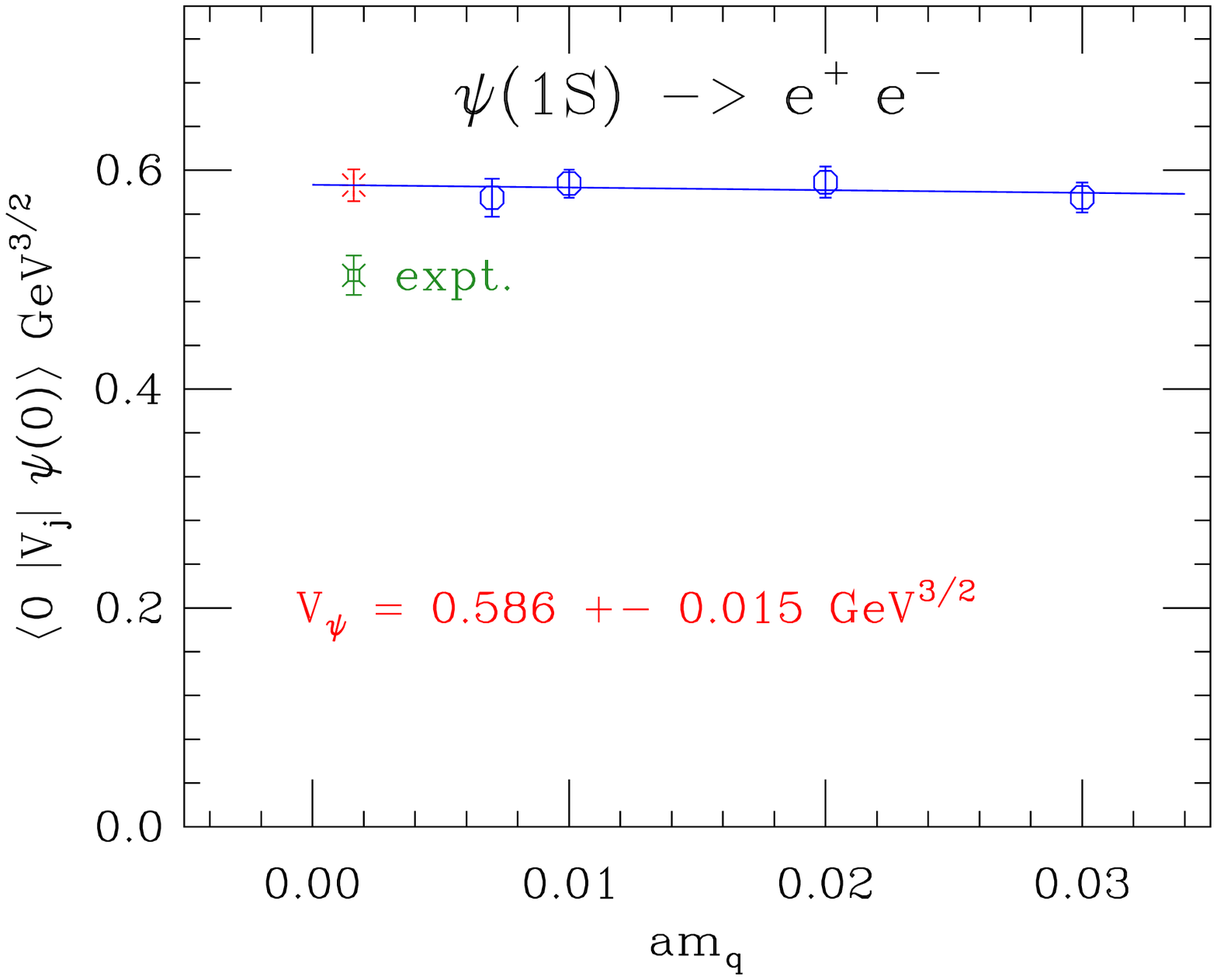}
\fi
\vspace{-5ex}
\caption{The annihilation matrix element for the decay
$\psi(\OneS)\to e^+ e^-$ as a function
of the mass of the two light sea quarks.
The extrapolated matrix element lies above
the value derived from the experimental leptonic
decay width.
}
\label{fig:leptonic}
\end{figure}}

\newcommand{\TABmilcLatts}{%
\begin{table}[htb]
\caption{The bare light sea quark mass and
coupling $\beta$
for the MILC three-flavor gluon ensembles.
The strange quark has mass $am_s=0.05$.
Lattice spacings are given in \GeV.
The $\TwoS\mbox{-}\OneS$ results are
from Reference~\cite{Wingate}.}
\label{table:milcLatts}
\begin{center}
\begin{tabular}{lllll}
$am_l$	&beta	&cfgs &$a^{-1}_{\psi}(\OneP\mbox{-}\OneS)$
&$a^{-1}_{\Upsilon}(\TwoS\mbox{-}\OneS)$ \\ \hline
$.007$	&$6.76$	&$403$ &1.55(3) &------ \\
$0.01$	&$6.76$ &$593$ &1.56(2) &1.59(2) \\
$0.02$	&$6.79$ &$460$ &1.59(3) &1.61(2) \\
$0.03$	&$6.81$ &$549$ &1.57(3) &1.60(3)
\end{tabular}\end{center}\end{table}}

\begin{document}

\begin{abstract}
We use the dynamical gluon configurations provided by the
MILC collaboration in a study of the charmonium spectrum 
and $\psi$ leptonic width. We examine sea quark effects
on mass splitting and on the leptonic decay matrix element
for light masses as low as $m_s/5$, while keeping the strange 
quark mass fixed and the lattice spacing nearly constant.

\vspace{1pc}
\end{abstract}

\maketitle

\section{INTRODUCTION}

Charmonium states below open flavor threshold are considered
gold-plated quantities in lattice QCD. They are almost
stable mesons, which should be accurately calculable in 
lattice QCD once realistic sea-quark effects are included 
in the simulations. A high precision study of the charmonium 
system in unquenched lattice QCD is interesting for several 
reasons. First, it provides us with an important test of the
lattice methods, because the methods used for charmonium
calculations are similar to those used for CKM determinations
in the $D$ meson system \cite{Mackenzie,Okamoto}.
Second, it allows us to test our improved actions which
are under development \cite{Oktay}, since different
splittings in the charmonium system are sensitive to
different correction operators in the action. For example,
the hyperfine splitting is very sensitive to 
$\bar\psi \sigma \cdot B \psi$, while the $\chi_c$ fine
structure is expected to sensitively depend on 
$\bar\psi \sigma \cdot(D \times E)\psi$. 
Finally, together with 2-loop perturbation theory, it yields 
precise determinations of $\alpha_s$ and the charm quark 
mass.

The $2+1$ dynamical gauge configurations
generated by the MILC collaboration contain realistic
sea quark effects, since they reach the chiral region
($m_l \geq m_s/5$) -- a necessary feature to control
chiral extrapolation errors. A first test of lattice QCD
calculations which use the MILC configurations was
presented in Ref.~\cite{Davies:2003ik}. For the first time
agreement (at the few \% level) with experiment was 
achieved for a variety of different physical systems, 
involving $b$, $c$, and light quarks. This comparison
includes our results for the 1P-1S splitting in charmonium
which are also presented here.

The work presented here continues our charmonium
study \cite{El-Khadra:zs,DiPierro:2002ta}.
Our companion study of the $D_s$ and $D$ meson 
spectra and weak decays is presented in 
Ref.~\cite{Mackenzie,Okamoto}.

\section{METHODS}

We are using the MILC collaboration
``Asqtad'' gluon ensembles \cite{Gottlieb,Bernard:2001av}. The Asqtad
action
has leading $\mathcal{O}(\alpha_s^2a^2)$ gluon uncertainties and
leading $\mathcal{O}(\alpha_s a^2)$ uncertainties
for the improved staggered sea quarks.

The gluon ensembles have
one flavor approximating
the strange quark and two equal-mass lighter flavors.
A matched set of gluon ensembles is available
having light masses in the range $m_s$ to $m_s/5$
and nearly constant lattice spacing
(see Table~\ref{table:milcLatts}).

Our charm quarks are $\mathcal{O}(a)$-improved Wilson
fermions in the Fermilab interpretation of heavy quarks.
The coefficient of the clover term  has the
tadpole-improved tree-level value.
The bare charm quark mass is tuned by demanding that the
$D_s$ meson kinetic mass equal the experimental value.

\section{CHARMONIUM SPECTRUM}

Fig.~\ref{fig:ccSpectrum} shows the overall picture of the
charmonium spectrum after sea quark extrapolations.
The zero of energy is taken to be the spin average
of the $\OneS$ masses. The lattice spacing is determined using
the $h_c(\OneP)$ splitting as input. This lattice spacing
is consistent, at the few percent level,
with other ways of setting the lattice spacing \cite{Davies:2003ik}.
We compare the charmonium $\OneP$-$\OneS$
and bottomonium $\TwoS$-$\OneS$ lattice spacings in
Table~\ref{table:milcLatts}.

\TABmilcLatts

\FIGccSpectrum

We extrapolate linearly in the light sea quark mass.
The mass dependence is mild:
linear terms are typically
of the same order of magnitude as their statistical error.
The
mass dependence for the hyperfine splitting,
shown in Fig.~\ref{fig:hyperfine},
illustrates a typical extrapolation.
Smaller statistical errors and more sea quark masses
would be needed to better resolve terms in the chiral
expansions.

The $\TwoS$ splittings shown in Fig.~\ref{fig:ccSpectrum} have
large errors.
Statistical uncertainties for these excited
states are $20-30$ times as large as uncertainties
in the ground states
with the same $J^{PC}$. Ground state and
radially excited state energies
are obtained from a single fit using Bayesian techniques.
We continue
to investigate ways to improve
the signal for excited states.

\subsection{Hyperfine splitting}

\FIGhyperfine

The hyperfine splitting in charmonium is a gold-plated
quantity, which must agree with experiment once all
known systematic errors are corrected.

As shown in Fig.~\ref{fig:hyperfine} we obtain a 
$\OneS$ hyperfine splitting of $97\pm2\;\MeV$, or in 
ratio to experiment, theory/expt$ = 0.82\pm0.02$.
A comparison to our previous quenched result (obtained
at similar lattice spacings), theory/expt$ = 0.6$,
shows that sea quarks have an appreciable effect
on this quantity. The remaining discrepancy is
likely due to having only a (tadpole improved)
tree-level estimate for the coefficient of the
$\sigma\cdot B$ term in our action. A one-loop 
calculation of this coefficient is in progress
\cite{Nobes}.

\section{$\mathbf{\psi(\OneS)}$ LEPTONIC WIDTH}

We determine the hadronic matrix element,
$V_\psi\equiv {\braOket{0}{V_j}{\psi}}$ for the 
leptonic width from the overlap coefficient of
the two-point function of the $\psi$ propagator
annihilated by the local vector current. At present 
our calculation does not include the $O(a)$ correction
for the vector current. 

For the spatial current renormalization, we use the 
formula $Z_{V_i} = \rho_{V_i} Z_{V_4}$,
and the nonperturbative result for $Z_{V_4}$ from 
Ref.~\cite{Okamoto}. Since the one-loop correction to
$\rho_{V_i}$ is currently unknown, we have
$\rho_{V_i} = 1$ in our calculation of $V_{\psi}$.
Hence, our results for this quantity are very 
preliminary.

The experimental measurement of the leptonic width
implies for the matrix element 
$V_\psi^{\rm expt}=0.504\pm0.018\,\GeV^{3/2}$.

\FIGleptonic

Fig.~\ref{fig:leptonic} shows our preliminary results 
for $V_\psi$; after sea quark extrapolation we find
$V_\psi^{\rm thy}=0.586\pm0.015\;\GeV^{3/2}$.
The statistics only uncertainty
is dominantly from the lattice spacing determination. We
find theory/expt$=1.16\pm0.04$, combining errors in quadrature.

\section*{ACKNOWLEDGMENTS}

We thank the MILC collaboration for their gauge configurations.
These calculations were done on the PC clusters deployed at
Fermilab under the DoE SciDAC program. This work was supported
in part by the DoE.

\end{document}

%% file: Macros.tex
\newcommand{\simgt}%
{\mathrel{\lower0.5ex\hbox{\ensuremath{\buildrel>\over\sim}}}}
\newcommand{\simlt}%
{\mathrel{\lower0.5ex\hbox{\ensuremath{\buildrel<\over\sim}}}}
\newcommand{\BarLongRightArrow}%
{\ensuremath{\mathrel\vert\joinrel\Longrightarrow}}

\newcommand{\GeV}{\textrm{GeV}}

\newcommand{\MeV}{\textrm{MeV}}

\newcommand{\fm}{\textrm{fm}}

\newcommand{\OneS}{\ensuremath{1S}}
\newcommand{\TwoS}{\ensuremath{2S}}
\newcommand{\OneP}{\ensuremath{1P}}

\newcommand{\braOket}[3]{%
\ensuremath{\left\langle{\,#1\,}\mid{\,#2\,}\mid{\,#3\,}\right\rangle}}

